\documentclass[a4paper,12pt]{article}

\usepackage{graphicx}
\usepackage{epstopdf} 
\usepackage[utf8]{inputenc}
\usepackage{cases}
\usepackage{epsfig}
\usepackage{amsfonts}
\usepackage{pdfpages}

\begin{document}

\begin{titlepage}
\vskip 2cm
\begin{center}
{\Large\bf  A method for solving nonlinear differential equations:  an application to  $\lambda\phi^4$ model}

 \vskip 10pt
{\bf  Danilo V. Ruy \\} \vskip 5pt
{\sl Instituto de Física Teórica-UNESP\\
Rua Dr Bento Teobaldo Ferraz 271, Bloco II, \\
01140-070, São Paulo, Brazil\\}  \vskip 2pt
\end{center}

\renewcommand{\abstractname}{Abstract}
\begin{abstract}
Recently, it has been great interest in the development of methods for solving nonlinear differential equations directly. Here, it is shown an algorithm based on Pad\'e approximants for solving nonlinear partial differential equations without requiring a one-dimensional reduction. This method is applied to the $\lambda\phi^4$ model in 4 dimensions and new solutions are obtained.

\end{abstract}

\bigskip


\vskip 3pt

\today
\end{titlepage}

\title{\first{A method for solving nonlinear differential equations: \\ an application to  $\lambda\phi^4$ model}}

\section{Introduction}

For many years, nonlinear differential equations have been an important topic of study and in many branches of knowledge. This interest has led to the development of many techniques through the last few years in order to obtain exact solutions without requiring further properties of the differential equation (for example \cite{Parkes,Malfliet,Kudryashov01,kudryashov,Parkes2,Vitanov1,Vitanov2,Vitanov3,Biswas1,Biswas2,He,Wu,Ma1,Ma2,Wang,Ebadi,Lee,Ruy}). In \cite{Ma1}, it was proposed the multiple Exp-function method for finding exact solutions of partial differential equation (PDE) without requiring a one-dimensional reduction. Although this method is very powerful, there are a large number of parameters to be determined.

Here, we present a simpler algorithm based on Pad\'e approximants and apply it to the classical equation of the $\lambda\phi^4$ model with $m\neq0$. There are several papers generalizing Pad\'e approximants for multivariate functions \cite{Chaffy,Chisholm,Cuyt,Guillaume00,Guillaume01,Guillaume,Levin}. Here, we focus on the homogeneous Pad\'e approximant, introduced in \cite{Cuyt} (see also \cite{Guillaume}).

The $\lambda \phi^4$ model is one of the simplest example of a renormalizable scalar field theory and it is defined by the Lagrangian density
\begin{equation}\label{lagrangian}
{\cal L}={1\over2}\partial_\mu\phi\partial^\mu\phi-{m^2\over 2}\phi^2-{\lambda\over 4}\phi^4,
\end{equation}
where we use the metric $\eta=(-+++)$ and the notation of repeated indices summed. The Euler-Lagrange equation, i. e
\[
\partial_\mu{\partial{\cal L}\over \partial\partial_\mu\phi}-{\partial{\cal L}\over \partial\phi}=0,
\]
of (\ref{lagrangian}) yields the classical equation of motion
\begin{equation}\label{intro}
-\partial_t^2\phi+\nabla^2\phi+m^2\phi+\lambda\phi^3=0.
\end{equation}
Static solutions of the two-dimensional $\lambda\phi^4$ model was already presented in \cite{Maia} and \cite{Dashen}. Here, we construct some solutions of equation (\ref{intro}) representing travelling waves and the scattering of two travelling waves.

The paper is organized in two main sections. In section \ref{section 2}, I show a pratical introduction to the homogeneous multivariate Pad\'e approximant and present a new approach for solving nonlinear PDEs. The section \ref{section 3} is devoted to the $\lambda\phi^4$ model.


\section{The method}\label{section 2}

\subsection{Homogeneous multivariate Pad\'e approximants}\label{explicacao Pade}

Consider a function $f({\bf z})=f(z_1,...,z_D)$ regular at origin and with Taylor expansion around ${\bf z}=0$ given by
\[
f({\bf z})=\sum_{{\bf J}=0}^\infty c_{{\bf J}}{\bf z}^{\bf J} \equiv \sum_{j_1=0}^\infty\sum_{j_2=0}^\infty ...\sum_{j_D=0}^\infty c_{j_1,j_2,...,j_D}\prod_{d=1}^D z_{d}^{j_d}.
\]

The homogeneous multivariate Pad\'e approximant consist in rearranging the coefficients such that we can use the Pad\'e approximant in one dimension. Through the map ${\bf z}\to \xi{\bf z}$, the Taylor Expansion can be rearranged as
\[
f({\xi \bf z})= \sum_{n=0}^{L+M}a_n({\bf z}) \xi^n+{\cal O}(\xi^{L+M+1})
\]
where
\[
a_n({\bf z})=\sum_{j_1=0}^{n}\sum_{j_2=0}^{n-j_1} ...\sum_{j_{D-1}=0}^{n-\sum_{r=1}^{D-2} j_{r}} c_{j_1,j_2,...,j_{D-1},n-\sum_{r=1}^{D-1} j_{r}}\biggl(\prod_{d=1}^{D-1} z_{d}^{j_d}\biggr) z_D^{n-\sum_{r=1}^{D-1} j_{r}}
\]

This rearrangement allows us to compute the univariat Pad\'e approximant of $f({\xi \bf z})$ on $\xi$, i. e.
\[
[L/M]_{\bf z}(\xi)\equiv{P_{{\bf z},L}(\xi)\over Q_{{\bf z},M}(\xi)}={\sum_{j=0}^L p_j({\bf z})\xi^j\over 1+\sum_{j=1}^M q_j({\bf z})\xi^j},
\]
where the coefficients $p_j({\bf z})$ and $q_j({\bf z})$ are determined such that the Pad\'e approximant agrees with $f({\xi \bf z})$ up to the degree $L+M$, i. e. $f(\xi {\bf z})=[L/M]_{\bf z}(\xi)+{\cal O}(\xi^{L+M+1})$. Thus we need to solve the following system of equations
\[
\sum_{r+s=j}a_r({\bf z})q_s({\bf z})-p_j({\bf z})=0,   \hspace{2 cm}   j=0,1,...,L+M.
\]

This system can be easily solved by a symbolic computation software. Concluding, the homogeneous Pad\'e approximant for f({\bf z}) is obtained by setting $f({\bf z})=f(\xi{\bf z})\biggl|_{\xi=1}$.


\subsection{The functional ansatz}\label{tecnica}

Consider now a system of $N_e$ equations with $N_e$ fields in D dimensions, i. e.
\[
E_k(x^\mu,\phi_i,\partial_\mu\phi_i,...;{\cal S}_0)=0,    \hspace{1 cm}   k=1,...,N_e
\]
where ${\cal S}_0$ is the space formed by the Cartesian product of the set of parameters, $\mu=1,...,D$ and $i=1,...,N_e$. Now, let us suppose at least one solution for this system can be expressed as  functionals of a set of functions $\rho=(\rho_1,...,\rho_{N_\rho})$, i.e.
\[
\phi_i(x^\mu)=\hat{\phi}_i(\rho_1,...,\rho_{N_\rho})  ,    \hspace{1 cm}   i=1,...,N_e
\]
where $\rho_k=\rho_k(x^\mu)$ for $k=1,...,N_\rho$.
Moreover, suppose the first derivative of all $\rho_k$ are known in terms of $\rho$, i. e.
\begin{equation}\label{derivadas primeiras}
\partial_\mu\rho_k={\cal F}_{\mu,k}(\rho_1,...,\rho_{N_\rho};{\cal S}_1),   \hspace{1 cm}   \mu=1,...,D   ,  \hspace{1 cm}   k=1,...,N_\rho,
\end{equation}
where ${\cal S}_1$  is the space formed by the Cartesian product of the sets of parameters introduced by  $\rho$. 
The choice of $\rho$ is the first ansatz of the algorithm and it yields the transformation
\begin{equation}\label{eq E}
E_k(x^\mu,\phi_i,\partial_\mu\phi_i,...;{\cal S}_0)=\hat{E}_k(\rho_k,\hat{\phi}_i,\partial_k\hat{\phi}_i,...;{\cal S}_0\times{\cal S}_1)=0,    \hspace{1 cm}   k=1,...,N_e,
\end{equation}
where $\hat{E}_k$ is polynomial or rational in $\rho_k$, $\hat{\phi}_i$ and its derivatives. Now, the system (\ref{eq E}) can be worked out as a system in $N_\rho$ dimensions. If at least one particular solution for the set of fields $\hat{\phi}_i$ are regular at origin, we can consider a multivariate Taylor expansion at $\rho=0$, i. e.
\begin{equation}\label{Taylor teorico}
\hat{\phi}_i(\rho)=\sum_{j_1=0}^\infty\sum_{j_2=0}^\infty ...\sum_{j_{N_\rho}=0}^\infty c_{i;j_1,j_2,...,j_{N_\rho}}\prod_{d=1}^{N_\rho} \rho_{d}^{j_d}  ,    \hspace{1 cm}   i=1,...,N_e
\end{equation}
where $c_{i;j_1,j_2,...,j_{N_\rho}}=c_{i;j_1,j_2,...,j_{N_\rho}}({\cal S}_0\times{\cal S}_1)$. Observe that the expansion (\ref{Taylor teorico}) can be drastically changed due to the particular combination of the parameters in the space formed by ${\cal S}_0\times{\cal S}_1$. Therefore, for obtaining  particular solutions, we can consider a set of constraints $\psi_i=\psi_i({\cal S}_0\times{\cal S}_1)=0$ acting on  $\rho$ and $\hat{E}_k(\rho_k,\hat{\phi}_i,\partial_k\hat{\phi}_i,...;{\cal S}_0\times{\cal S}_1)=0$. When a constraint $\psi$ is considered, the notation $\bar{\rho}\equiv \rho|_{\psi=0}$ will be employed.

Let us call ${\cal S}_2$  the space formed by the Cartesian product of all undetermined $c_{i;j_1,j_2,...,j_{N_\rho}}$ and define ${\cal S}={\cal S}_0\times{\cal S}_1\times{\cal S}_2$. Now, we can use the homogeneous multivariate Pad\'e approximant in $\hat{\phi}_i(\rho)$. Mapping $\rho\to\xi\rho$, we can rearrange the expansion (\ref{Taylor teorico}) as
\[
\hat{\phi}_i(\xi\rho)= \sum_{n=0}^{L_i+M_i}a_{i;n}(\rho) \xi^n+{\cal O}(\xi^{L_i+M_i+1})  ,    \hspace{1 cm}   i=1,...,N_e  ,
\]
\[
a_{i;n}(\rho)=\sum_{j_1=0}^{n}\sum_{j_2=0}^{n-j_1} ...\sum_{j_{{N_\rho}-1}=0}^{n-\sum_{r=1}^{{N_\rho}-2} j_{r}} c_{i;j_1,j_2,...,j_{{N_\rho}-1},n-\sum_{r=1}^{{N_\rho}-1} j_{r}}\biggl(\prod_{d=1}^{{N_\rho}-1} \rho_{d}^{j_d}\biggr) \rho_{N_\rho}^{n-\sum_{r=1}^{{N_\rho}-1} j_{r}}
\]

Thus, we can apply the univariate Pad\'e approximant on $\xi$ in order to obtain an approximation of the solution, i. e.
\begin{equation}\label{aproximado}
\hat{\phi}_i(\xi \rho )={P_{\rho,L_i}(\xi;{\cal S})\over Q_{\rho,M_i}(\xi;{\cal S})}+{\cal O}(\xi^{L_i+M_i+1}).
\end{equation}

Finally, we can apply the second ansatz. Let us assume that there is a particular subset $\hat{{\cal S}}\subset{\cal S}$, such that expression (\ref{aproximado}) yields an exact solution when $\xi=1$, i. e.  
\begin{equation}\label{sol exata}
\hat{\phi}_i(\rho)={P_{\rho,L_i}(\xi;\hat{{\cal S}})\over Q_{\rho,M_i}(\xi;\hat{{\cal S}})}\biggl|_{\xi=1}.
\end{equation}

This idea was used in \cite{Ruy} for the simpler case when $D=N_e=N_\rho=1$ and $\rho_1=z$ (where $z$ was the one-dimensional variable). In order to determine $\hat{{\cal S}}$, let us substitute (\ref{sol exata}) in (\ref{eq E}). This yields
\[
\hat{E}_k(\rho_k,\hat{\phi}_i,\partial_k\hat{\phi}_i,...;{\cal S}_0\times{\cal S}_1)={\sum_{n=0}^{\Lambda}\hat{E}_{k;n}(\hat{S})  \over D_k(\hat{S}) }=0,
\]
\begin{equation}
\hat{E}_{k;n}(\hat{S})=\sum_{j_1=0}^{n}\sum_{j_2=0}^{n-j_1} ...\sum_{j_{{N_\rho}-1}=0}^{n-\sum_{r=1}^{{N_\rho}-2} j_{r}} \tilde{E}_{k;j_1,j_2,...,j_{{N_\rho}-1},n-\sum_{r=1}^{{N_\rho}-1} j_{r}}\biggl(\prod_{d=1}^{{N_\rho}-1} \rho_{d}^{j_d}\biggr) \rho_{N_\rho}^{n-\sum_{r=1}^{{N_\rho}-1} j_{r}}.
\end{equation}
where $\Lambda$ is determined by the choices of $L_i$, $M_i$, the set $\rho$ and the differential equation under consideration. Hence, in order to determine all elements of $\hat{{\cal S}}$, we need to solve the following algebraic system:
\begin{subequations}
\begin{eqnarray}\label{cond teorico}
\tilde{E}_{k;j_1,j_2,...,j_{{N_\rho}}}(\hat{S}) = 0  ,  &&       k=1,...,N_e  ,   \hspace{0.4 cm}  n=0,...,\Lambda  ,   \hspace{0.4 cm} j_l=0,...,n-\sum_{r=1}^{l-1}j_r  \\
 D_k(\hat{S}) \neq 0  ,   &&    k=1,...,N_e 
\end{eqnarray}
\end{subequations}


Step (\ref{cond teorico}) may require a huge computational power for some models, but it yields a system of algebraic equations smaller than the one we should solve by using the multiple Exp-function method \cite{Ma1}.

\section{An application to the $\lambda\phi^4$ theory in 4 dimensions}\label{section 3}

Here, we apply the algorithm presented in section \ref{tecnica} for the $\lambda\phi^4$ theory, i. e.
\begin{equation}\label{lambda phi 4}
-\partial_t^2\phi+\nabla^2\phi+m^2\phi+\lambda\phi^3=0 ,
\end{equation}
by using two different functional ansatz:
\begin{eqnarray}
(i) && \phi(x^\mu)=\hat{\phi}(\rho_1),   \hspace{1.05 cm}  \rho_1=e^{i(k_{1,0}t+k_{1,1}x+k_{1,2}y+k_{1,3}z)}  \\
&& \nonumber  \\
(ii) && \phi(x^\mu)=\hat{\phi}(\rho_1,\rho_2),   \hspace{0.5 cm}  \rho_1=e^{i(k_{1,0}t+k_{1,1}x+k_{1,2}y+k_{1,3}z)},   \hspace{0.5 cm}  \rho_2=e^{i(k_{2,0}t+k_{2,1}x+k_{2,2}y+k_{2,3}z)}  \nonumber \\
\end{eqnarray}

For simplicity, let us use the notation ${\bf k_j}=(k_{j,1},k_{j,2},k_{j,3})$, where ${\bf k_i}.{\bf k_j}=k_{i,1}k_{j,1}+k_{i,2}k_{j,2}+k_{i,3}k_{j,3}$.

\subsection{Ansatz (i)}

First, consider the ansatz (i). Obviously, this set for $\rho$ satisfies condition (\ref{derivadas primeiras}), namely
\[
\partial_t\rho_1=i k_{1,0}\rho_1,   \hspace{1 cm}  \partial_x\rho_1=i k_{1,1}\rho_1,   \hspace{1 cm}  \partial_y\rho_1=i k_{1,2}\rho_1,   \hspace{1 cm} \partial_z\rho_1=i k_{1,3}\rho_1,
\]
and yields the equation
\begin{equation}\label{lambda phi 4 i}
\hat{E}(\rho_1,\hat{\phi},\partial_{\rho_1}\hat{\phi},\partial^2_{\rho_1}\hat{\phi};{\cal S}_0\times{\cal S}_1)\equiv(k_{1,0}^2-{\bf k_1}^2)(\rho_1^2\partial^2_{\rho_1}\hat{\phi}+\rho_1\partial_{\rho_1}\hat{\phi})+m^2\hat{\phi}+\lambda\hat{\phi}^3=0.
\end{equation}

The first element of the Taylor expansion of $\phi$ can be $c_0=0$ or $c_0={i\mu m\over\sqrt{\lambda}}$ where $\mu=\pm1$. Without imposing any constraint $\psi$, this expansion yields two trivial solutions, namely,
\begin{equation}\label{sol contants}
\hat{\phi}=0   \hspace{1 cm}   \textnormal{and}   \hspace{1 cm}  \hat{\phi}={i\mu m\over\sqrt{\lambda}},   \hspace{0.7 cm}   \mu=\pm1.
\end{equation}
However, a convenient choice for $\psi$ leads us to a more interesting expansion. If we impose
\begin{equation}\label{psi v1 - 1}
\psi=-k_{1,0}^2+{\bf k_1}^2-m^2=0
\end{equation}
on equation (\ref{lambda phi 4 i}) and the set $\rho$  by eliminating $k_{1,0}$, we get
\begin{eqnarray}\label{lambda phi 4 i psi1}
&& -m^2(\bar{\rho}_1^2\partial^2_{\bar{\rho}_1}\hat{\phi}+\bar{\rho}_1\partial_{\bar{\rho}_1}\hat{\phi})+m^2\hat{\phi}+\lambda\hat{\phi}^3=0,  
\end{eqnarray}
where $\bar{\rho}_1=e^{i(\nu\sqrt{({\bf k_1}^2-m^2)}t+k_{1,1}x+k_{1,2}y+k_{1,3}z)}$ and $\nu=\pm1$. The expansion of $\hat{\phi}$ starting with $c_0=0$ then has the form
\begin{equation}\label{expansao N1 v1}
\hat{\phi}=(c_1 \bar{\rho}_1)\xi+\biggl({c_1^3 \lambda\over 8m^2}\bar{\rho}^3\biggr)\xi^3+\biggl({c_1^5 \lambda^2\over 64m^4}\bar{\rho}^5\biggr)\xi^5+\biggl({c_1^7 \lambda^3\over 512m^6}\bar{\rho}^7\biggr)\xi^7+...,  \hspace{1 cm}  m\neq 0
\end{equation}
while the expansion starting with $c_0={i\mu m\over\sqrt{\lambda}}$ truncate at the constant term.
By employing the Pad\'e approximant $[1/1]_{(\rho_1)}(\xi)|_{\xi=1}$  of expansion (\ref{expansao N1 v1}), ansatz (\ref{sol exata}) yields
\begin{equation}\label{sol v1 -1/1}
\hat{\phi}=c_1 \bar{\rho}_1,     \hspace{1 cm}
\end{equation}
whose substitution into equation (\ref{lambda phi 4 i psi1}) leads us to the conditions:
\begin{eqnarray*}
\tilde{E}_{1;3}(\hat{S}) &=& c_1^3\lambda=0  ,  \\
 D_1(\hat{S}) &=& 1  \neq 0.
\end{eqnarray*}

Here, we see that the condition for (\ref{sol v1 -1/1}) to be a solution of  (\ref{lambda phi 4 i psi1})  is $c_1=0$ or $\lambda=0$, which represents the vacuum solution and the Klein-Gordon limit respectively. Now, let us consider the Pad\'e approximant $[2/2]_{(\rho_1)}(\xi)|_{\xi=1}$. In this case, the ansatz (\ref{sol exata}) yields
\begin{equation}\label{sol v1 -2/2}
\hat{\phi}={8c_1 m^2 \bar{\rho}_1\over 8m^2- c_1^2\lambda \bar{\rho}_1^2}
\end{equation}

By substituting (\ref{sol v1 -2/2}) into (\ref{lambda phi 4 i psi1}), we can check that expression (\ref{sol v1 -2/2}) already is an exact solution without requiring any further conditions. We also obtain (\ref{sol v1 -2/2})  if we use the ansatz $[3/3]_{(\rho_1)}(\xi)|_{\xi=1}$, $[4/4]_{(\rho_1)}(\xi)|_{\xi=1}$ or $[5/5]_{(\rho_1)}(\xi)|_{\xi=1}$.

Now, consider the constraint
\[
\psi=-k_{1,0}^2+{\bf k_1}^2+2m^2=0.
\]
By eliminating $k_{1,0}$, equation (\ref{lambda phi 4 i}) yields
\begin{eqnarray}\label{lambda phi 4 i psi1}
&& 2m^2(\bar{\rho}_1^2\partial^2_{\bar{\rho}_1}\hat{\phi}+\bar{\rho}_1\partial_{\bar{\rho}_1}\hat{\phi})+m^2\hat{\phi}+\lambda\hat{\phi}^3=0,  
\end{eqnarray}
where $\bar{\rho}_1=e^{i(\nu\sqrt{({\bf k_1}^2+2m^2)}t+k_{1,1}x+k_{1,2}y+k_{1,3}z)}$ for $\nu=\pm1$. Using this constraint, the expansion of $\hat{\phi}$ starting with $c_0=0$ truncates at the constant term, while the expansion starting with $c_0={i\mu m\over \sqrt{\lambda}}$ has the form
\begin{equation}\label{expansao N1 v2}
\hat{\phi}={i\mu m\over \sqrt{\lambda}}+(c_1\bar{\rho}_1)\xi-\biggl({i\mu c_1^2\sqrt{\lambda}\over 2m}\bar{\rho}_1^2\biggr)\xi^2-\biggl({c_1^3\lambda\over 4m^2}\bar{\rho}_1^3\biggr)\xi^3+\biggl({i\mu c_1^4\lambda^{3/2}\over 8m^3}\bar{\rho}_1^4\biggr)\xi^4+... \hspace{1 cm}  m\neq 0.
\end{equation}

By employing the Pad\'e approximant $[1/1]_{(\rho_1)}(\xi)|_{\xi=1}$ and $[2/2]_{(\rho_1)}(\xi)|_{\xi=1}$  of expansion (\ref{expansao N1 v2}), ansatz (\ref{sol exata}) yields
\begin{eqnarray}
\hat{\phi}=[1/1]_{(\rho_1)}(\xi)|_{\xi=1}&=&{m(c_1 \sqrt{\lambda}\bar{\rho}_1+2i\mu m)\over \sqrt{\lambda}(2m+i\mu c_1\sqrt{\lambda}\bar{\rho}_1)},   \label{sol v2 -1/1}\\
\hat{\phi}=[2/2]_{(\rho_1)}(\xi)|_{\xi=1}&=&{4c_1\sqrt{\lambda}m^2 \bar{\rho}_1+i\mu(4m^3-c_1^2\lambda m \bar{\rho}_1^2)\over \sqrt{\lambda}(4m^2+c_1^2\lambda\bar{\rho}_1^2)}  \label{sol v2 -2/2}.
\end{eqnarray}
Both (\ref{sol v2 -1/1}) and (\ref{sol v2 -2/2}) are exact solutions without requiring further conditions on the space of constants ${\cal S}$.  The solution (\ref{sol v2 -2/2}) is also obtained if we consider the ansatz $[3/3]_{(\rho_1)}(\xi)|_{\xi=1}$, $[4/4]_{(\rho_1)}(\xi)|_{\xi=1}$ or $[5/5]_{(\rho_1)}(\xi)|_{\xi=1}$.

\subsection{Ansatz (ii)}

Now, consider the ansatz (ii). Clearly, this set for $\rho$ also satisfies condition (\ref{derivadas primeiras}), namely
\begin{eqnarray*}
&&\partial_t\rho_1=i k_{1,0}\rho_1,   \hspace{1 cm}  \partial_x\rho_1=i k_{1,1}\rho_1,   \hspace{1 cm}  \partial_y\rho_1=i k_{1,2}\rho_1,   \hspace{1 cm} \partial_z\rho_1=i k_{1,3}\rho_1,  \\
&&\partial_t\rho_2=i k_{2,0}\rho_2,   \hspace{1 cm}  \partial_x\rho_2=i k_{2,1}\rho_2,   \hspace{1 cm}  \partial_y\rho_2=i k_{2,2}\rho_2,   \hspace{1 cm} \partial_z\rho_2=i k_{2,3}\rho_2, 
\end{eqnarray*}
and yields the equation
\begin{eqnarray}\label{lambda phi 4 ii}
\hat{E}(\rho_1,\rho_2,\hat{\phi},,...;{\cal S}_0\times{\cal S}_1) &\equiv& (k_{1,0}^2-{\bf k_1}^2)(\rho_1^2\partial^2_{\rho_1}\hat{\phi}+\rho_1\partial_{\rho_1}\hat{\phi})+ (k_{2,0}^2-{\bf k_2}^2)(\rho_2^2\partial^2_{\rho_2}\hat{\phi}  \nonumber \\
&+& \rho_2\partial_{\rho_2}\hat{\phi})+ 2(k_{1,0}k_{2,0}-{\bf k_1}.{\bf k_2})\rho_1\rho_2\partial_{\rho_1}\partial_{\rho_2}\hat{\phi}+m^2\hat{\phi}  \nonumber \\
&+& \lambda\hat{\phi}^3=0.
\end{eqnarray}

Without imposing any constraint $\phi$, this functional ansatz yields solution (\ref{sol contants}). In addition, if we impose the constraint
\begin{equation}\label{constraint 1 - v1}
\psi_1=-k_{1,0}^2+{\bf k_1}^2-m^2=0
\end{equation}
or 
\begin{equation}\label{constraint 2 - v1}
\psi_2=-k_{1,0}^2+{\bf k_2}^2-m^2=0
\end{equation}
we obtain solution (\ref{sol v1 -2/2}) up to a redefinition of an arbitrary constant. However, if we employ constraints (\ref{constraint 1 - v1}) and (\ref{constraint 2 - v1}) and eliminate $k_{1,0}$ and $k_{2,0}$, equation (\ref{lambda phi 4 ii}) yields
\begin{equation}\label{reduzida (ii) 1}
-m^2(\rho_1^2\partial^2_{\rho_1}\hat{\phi}+\rho_1\partial_{\rho_1}\hat{\phi}+\rho_2^2\partial^2_{\rho_2}\hat{\phi}+ \rho_2\partial_{\rho_2}\hat{\phi})+ 2(k_{1,0}k_{2,0}-{\bf k_1}.{\bf k_2})\rho_1\rho_2\partial_{\rho_1}\partial_{\rho_2}\hat{\phi}+m^2\hat{\phi}+\lambda\hat{\phi}^3=0 
\end{equation}
%
%
Observe that we did not eliminate the linear terms of $k_{1,0}$ and $k_{2,0}$ at this stage in order to avoid mistakes with the sign of the root square. The multivariate Taylor expansion of (\ref{reduzida (ii) 1}) yields
\footnotesize
\begin{equation}\label{expansao N2 v1}
\hat{\phi}=(c_{1,0} \bar{\rho}_1+c_{0,1} \bar{\rho}_2)\xi+\Biggl({\lambda(c_{1,0}^3\bar{\rho}_1^3+c_{0,1}^3\bar{\rho}_2^3)\over 8m^2} - {3\lambda(c_{1,0}^2c_{0,1}\bar{\rho}_1^2\bar{\rho}_2+c_{1,0}c_{0,1}^2\bar{\rho}_1\bar{\rho}_2^2)\over 4(k_{1,0}k_{2,0}-{\bf k_1}.{\bf k_2}-m^2)}\Biggr)\xi^3+...,
\end{equation}
\normalsize
for $m\neq0$ and $k_{1,0}k_{2,0}-{\bf k_1}.{\bf k_2}-m^2\neq 0$. 

The Pad\'e approximant $[1/1]_{(\rho_1,\rho_2)}(\xi)|_{\xi=1}$  of expansion (\ref{expansao N2 v1}) yield the ansatz
\begin{equation}\label{sol N2 v1 1/1}
\hat{\phi}=c_{1,0}\bar{\rho}_1+c_{0,1}\bar{\rho}_2,
\end{equation}
such that the conditions for (\ref{sol N2 v1 1/1}) to become an exact solution are
\begin{eqnarray*}
\tilde{E}_{1;3,0}(\hat{S}) &=& c_{1,0}^3\lambda=0  ,  \\
\tilde{E}_{1;2,1}(\hat{S}) &=& 3c_{1,0}^2c_{0,1}\lambda=0  ,  \\
\tilde{E}_{1;1,2}(\hat{S}) &=& 3c_{1,0}c_{0,1}^2\lambda=0  ,  \\
\tilde{E}_{1;0,3}(\hat{S}) &=& c_{0,1}^3\lambda=0  ,  \\
 D_1(\hat{S}) &=& 1  \neq 0.
\end{eqnarray*}

Hence, the only possibility for (\ref{sol N2 v1 1/1}) to be a solution of (\ref{lambda phi 4 ii}) is in the Klein-Gordon limit, i. e. $\lambda=0$. Employing the ansatz $[2/2]_{(\rho_1,\rho_2)}(\xi)|_{\xi=1}$, we obtain
\begin{equation}\label{sol N2 v1 2/2}
\hat{\phi}={8m^2(-k_{1,0}k_{2,0}+{\bf k_1}.{\bf k_2}+m^2)(c_{1,0}\bar{\rho}_1+c_{0,1}\bar{\rho}_2)\over (k_{1,0}k_{2,0}-{\bf k_1}.{\bf k_2}-m^2)(\lambda(c_{1,0}^2\bar{\rho}_1^2-c_{1,0}c_{0,1}\bar{\rho}_1\bar{\rho}_2+c_{0,1}^2\bar{\rho}_2^2)-8m^2) - 6m^2c_{1,0}c_{0,1}\lambda\bar{\rho}_1\bar{\rho}_2}  
\end{equation}
By substituting ansatz (\ref{sol N2 v1 2/2}) into (\ref{lambda phi 4 ii}), we obtain the conditions
\begin{eqnarray*}
\tilde{E}_{1;4,1}(\hat{S}) &=& 64c_{1,0}^4c_{0,1}\lambda^2m^2(k_{1,0}k_{2,0}-{\bf k_1}.{\bf k_2}-m^2)^2(k_{1,0}k_{2,0}-{\bf k_1}.{\bf k_2}+m^2) \\
  && (k_{1,0}k_{2,0}-{\bf k_1}.{\bf k_2}+2m^2)=0  ,  \\
\tilde{E}_{1;3,2}(\hat{S}) &=& -96c_{1,0}^3c_{0,1}^2\lambda^2m^2(k_{1,0}k_{2,0}-{\bf k_1}.{\bf k_2}-m^2)^3 (k_{1,0}k_{2,0}-{\bf k_1}.{\bf k_2}+m^2)=0  ,  \\
\tilde{E}_{1;2,3}(\hat{S}) &=& -96c_{1,0}^2c_{0,1}^3\lambda^2m^2(k_{1,0}k_{2,0}-{\bf k_1}.{\bf k_2}-m^2)^3 (k_{1,0}k_{2,0}-{\bf k_1}.{\bf k_2}+m^2)=0  ,  \\
\tilde{E}_{1;1,4}(\hat{S}) &=& 64c_{1,0}c_{0,1}^4\lambda^2m^2(k_{1,0}k_{2,0}-{\bf k_1}.{\bf k_2}-m^2)^2(k_{1,0}k_{2,0}-{\bf k_1}.{\bf k_2}+m^2) \\
  && (k_{1,0}k_{2,0}-{\bf k_1}.{\bf k_2}+2m^2)=0  ,  \\  
 D_1(\hat{S}) &=& [(k_{1,0}k_{2,0}-{\bf k_1}.{\bf k_2}-m^2)(c_{1,0}^2\lambda\bar{\rho}_1^2-c_{1,0}c_{0,1}\lambda\bar{\rho}_1\bar{\rho}_2+c_{0,1}^2\lambda\bar{\rho}_2^2-8m^2) \\
  && -6m^2c_{1,0}c_{0,1}\lambda\bar{\rho}_1\bar{\rho}_2]^3  \neq 0.
\end{eqnarray*}

The system of conditions above has four possibilities for solutions, namely $c_{1,0}=0$, $c_{0,1}=0$, $\lambda=0$ or
\begin{equation}\label{cond N2}
k_{1,0}k_{2,0}-{\bf k_1}.{\bf k_2}+m^2=0.
\end{equation}
It is easy to see that conditions $c_{1,0}=0$ and $c_{0,1}=0$ yield the same solution (\ref{sol v1 -2/2}) up to a redefinition of the arbitrary constants, while the condition $\lambda=0$ yields the solution (\ref{sol N2 v1 1/1}).
We can also use condition (\ref{cond N2}) and the constraints (\ref{constraint 1 - v1}) and (\ref{constraint 2 - v1}) for eliminating three constants, for example
\footnotesize
\begin{eqnarray}
&& k_{1,0}={\nu_1k_{2,1}\sqrt{m^2\sum_{j=2}^3(k_{1,j}-k_{2,j})^2-(k_{1,3}k_{2,2}-k_{1,2}k_{2,3})^2}+\nu_2\sqrt{{\bf k_2}^2-m^2}(\sum_{j=2}^3k_{1,j}k_{2,j}-m^2)  \over \sum_{j=2}^3 k_{2,j}^2-m^2  }  \nonumber \\
&&  \label{cont v1-1}   \\
&& k_{2,0}=\nu_2\sqrt{{\bf k_2}^2-m^2} \label{cont v1-2} \\
&& k_{1,1}={k_{2,1}(\sum_{j=2}^3 k_{1,j}k_{2,j}-m^2)+\nu_1\nu_2\sqrt{({\bf k_2}^2-m^2)[m^2\sum_{j=2}^3(k_{1,j}-k_{2,j})^2-(k_{1,3}k_{2,2}-k_{1,2}k_{2,3})^2]}\over \sum_{j=2}^3 k_{2,j}^2-m^2  } \nonumber \\
&& \label{cont v1-3}
\end{eqnarray}
\normalsize
where $\nu_1=\pm1$ and $\nu_2=\pm1$. Therefore, expression (\ref{sol N2 v1 2/2}) is a solution of (\ref{lambda phi 4}) with $\bar{\rho}_1=e^{i(k_{1,0}t+k_{1,1}x+k_{1,2}y+k_{1,3}z)}$,  $\bar{\rho}_2=e^{i(k_{2,0}t+k_{2,1}x+k_{2,2}y+k_{2,3}z)}$ and the constants  (\ref{cont v1-1}), (\ref{cont v1-2}) and (\ref{cont v1-3}).

Now, let us consider equation (\ref{lambda phi 4 ii}) and the set $\rho$ with the constraints
\begin{eqnarray}
\psi_1&=&-k_{1,0}^2+{\bf k_1}^2+2m^2=0,    \label{contraint N2 v2 1}    \\
\psi_2&=&-k_{2,0}^2+{\bf k_2}^2+2m^2=0.    \label{contraint N2 v2 2} .
\end{eqnarray}
If we consider only one of these constants, we will obtain the solution (\ref{sol v2 -1/1}) again, up to a redefinition of the arbitrary constants. However, if both constraints are considered we can eliminate $k_{1,0}$, $k_{1,0}$, such that (\ref{lambda phi 4 ii}) yields
\begin{equation}\label{reduzida (ii) 2}
2m^2(\rho_1^2\partial^2_{\rho_1}\hat{\phi}+\rho_1\partial_{\rho_1}\hat{\phi}+\rho_2^2\partial^2_{\rho_2}\hat{\phi}+ \rho_2\partial_{\rho_2}\hat{\phi})+ 2(k_{1,0}k_{2,0}-{\bf k_1}.{\bf k_2})\rho_1\rho_2\partial_{\rho_1}\partial_{\rho_2}\hat{\phi}+m^2\hat{\phi}+\lambda\hat{\phi}^3=0. 
\end{equation}
%
Observe that we did not substitute linear terms of $k_{1,0}$ and $k_{2,0}$ again. The multivariate Taylor expansion of (\ref{reduzida (ii) 2}) yields
\footnotesize
\begin{equation}\label{expansao N2 v2}
\hat{\phi}={i\mu m\over \sqrt{\lambda}}+(c_{1,0} \bar{\rho}_1+c_{0,1} \bar{\rho}_2)\xi-\Biggl({i\mu\sqrt{\lambda}(c_{1,0}^2\bar{\rho}_1^2+c_{0,1}^2\bar{\rho}_2^2)\over 2m} + {3i\mu m\lambda c_{1,0}c_{0,1}\bar{\rho}_1\bar{\rho}_2\over (k_{1,0}k_{2,0}-{\bf k_1}.{\bf k_2}+m^2)}\Biggr)\xi^2+...,
\end{equation}
\normalsize
for $m\neq0$, $\lambda\neq0$ and $k_{1,0}k_{2,0}-{\bf k_1}.{\bf k_2}+m^2\neq 0$. The Pad\'e approximant  $[1/1]_{(\rho_1,\rho_2)}(\xi)|_{\xi=1}$ of (\ref{expansao N2 v2}) yields the ansatz
\begin{eqnarray}
\hat{\phi}&=& m\biggl((k_{1,0}k_{2,0}-{\bf k_1}.{\bf k_2}+m^2)(\sqrt{\lambda}(c_{1,0}^2\bar{\rho}_1^2+c_{0,1}^2\bar{\rho}_2^2+4c_{1,0}c_{0,1}\bar{\rho}_1\bar{\rho}_2)+2im\mu(c_{1,0}\bar{\rho}_1+c_{0,1}\bar{\rho}_2)) \nonumber \\
&-&6 m^2\sqrt{\lambda} c_{1,0}c_{0,1}\bar{\rho}_1\bar{\rho}_2  \biggr) \biggl/ \biggl((k_{1,0}k_{2,0}-{\bf k_1}.{\bf k_2}+m^2)(i\mu\lambda(c_{1,0}^2\bar{\rho}_1^2+c_{0,1}^2\bar{\rho}_2^2)+2m\sqrt{\lambda}(c_{1,0}\bar{\rho}_1  \nonumber \\
&+&c_{0,1}\bar{\rho}_2))+6i\mu\lambda m^2 c_{1,0}c_{0,1}\bar{\rho}_1\bar{\rho}_2\biggr)  ,   \label{sol N2 v2 1/1}
\end{eqnarray}
and the conditions for (\ref{sol N2 v2 1/1}) to become an exact solution are
\begin{eqnarray*}
\tilde{E}_{1;5,1}(\hat{S}) &=& 8c_{1,0}^5c_{0,1}\lambda^{3/2} m(k_{1,0}k_{2,0}-{\bf k_1}.{\bf k_2}-m^2)(k_{1,0}k_{2,0}-{\bf k_1}.{\bf k_2}+m^2)^2  \\
&&(k_{1,0}k_{2,0}-{\bf k_1}.{\bf k_2}-2m^2)=0  \\
\tilde{E}_{1;4,2}(\hat{S}) &=& -96c_{1,0}^4c_{0,1}^2\lambda^{3/2} m^5(k_{1,0}k_{2,0}-{\bf k_1}.{\bf k_2}+m^2) (k_{1,0}k_{2,0}-{\bf k_1}.{\bf k_2}-2m^2)=0 \\
\tilde{E}_{1;3,3}(\hat{S}) &=& 16c_{1,0}^3c_{0,1}^3\lambda^{3/2} m(k_{1,0}k_{2,0}-{\bf k_1}.{\bf k_2}+m^2)(k_{1,0}k_{2,0}-{\bf k_1}.{\bf k_2}-2m^2)[8m^2(k_{1,0}k_{2,0} \\
 &&-{\bf k_1}.{\bf k_2})-3((k_{1,0}k_{2,0}-{\bf k_1}.{\bf k_2}))^2+5m^4]=0  \\
\tilde{E}_{1;2,4}(\hat{S}) &=& -96c_{1,0}^2c_{0,1}^4\lambda^{3/2} m^5(k_{1,0}k_{2,0}-{\bf k_1}.{\bf k_2}+m^2)(k_{1,0}k_{2,0}-{\bf k_1}.{\bf k_2}-2m^2) \\
\tilde{E}_{1;1,5}(\hat{S}) &=& 8c_{1,0}c_{0,1}^5\lambda^{3/2} m(k_{1,0}k_{2,0}-{\bf k_1}.{\bf k_2}-m^2)(k_{1,0}k_{2,0}-{\bf k_1}.{\bf k_2}+m^2)^2 \\
 &&(k_{1,0}k_{2,0}-{\bf k_1}.{\bf k_2}-2m^2)=0  \\
D_1(\hat{S}) &=& [(k_{1,0}k_{2,0}-{\bf k_1}.{\bf k_2}+m^2)(i\mu\lambda(c_{1,0}^2\lambda\bar{\rho}_1^2+c_{0,1}^2\lambda\bar{\rho}_2^2)+2m\sqrt{\lambda}(c_{1,0}\bar{\rho}_1+c_{0,1}\bar{\rho}_2) \\
&+& 6i\mu\lambda m^2 c_{1,0}c_{0,1}\bar{\rho}_1\bar{\rho}_2\biggr)] \neq 0.
\end{eqnarray*}

The system for the above conditions above has three possibilities of solutions, namely $c_{1,0}=0$, $c_{0,1}=0$, or
\begin{equation}\label{cond N2 v2}
k_{1,0}k_{2,0}-{\bf k_1}.{\bf k_2}-2m^2=0.
\end{equation}
The conditions $c_{1,0}=0$ and $c_{0,1}=0$ yield the solution (\ref{sol v2 -1/1}) up to a redefinition of the arbitrary constants, while condition (\ref{cond N2 v2}) and the constraints (\ref{contraint N2 v2 1}) and (\ref{contraint N2 v2 2}) eliminate three constants, for example,
\footnotesize
\begin{eqnarray}
&& k_{1,0}={\nu_1k_{2,1}\sqrt{-2m^2\sum_{j=2}^3(k_{1,j}-k_{2,j})^2-(k_{1,3}k_{2,2}-k_{1,2}k_{2,3})^2}+\nu_2\sqrt{{\bf k_2}^2+2m^2}(\sum_{j=2}^3k_{1,j}k_{2,j}+2m^2)  \over \sum_{j=2}^3 k_{2,j}^2+2m^2  }  \nonumber \\
&&  \label{cont v2-1}   \\
&& k_{2,0}=\nu_2\sqrt{{\bf k_2}^2+2m^2} \label{cont v2-2} \\
&& k_{1,1}={k_{2,1}(\sum_{j=2}^3 k_{1,j}k_{2,j}+2m^2)+\nu_1\nu_2\sqrt{({\bf k_2}^2+2m^2)[-2m^2\sum_{j=2}^3(k_{1,j}-k_{2,j})^2-(k_{1,3}k_{2,2}-k_{1,2}k_{2,3})^2]}\over \sum_{j=2}^3 k_{2,j}^2+2m^2  } \nonumber \\
&& \label{cont v2-3}
\end{eqnarray}
\normalsize
where $\nu_1=\pm1$ and $\nu_2=\pm1$. Therefore, the expression (\ref{sol N2 v2 1/1}) is a solution of the $\lambda\phi^4$ model with $\bar{\rho}_1=e^{i(k_{1,0}t+k_{1,1}x+k_{1,2}y+k_{1,3}z)}$,  $\bar{\rho}_2=e^{i(k_{2,0}t+k_{2,1}x+k_{2,2}y+k_{2,3}z)}$ and the constants  (\ref{cont v2-1}), (\ref{cont v2-2}) and (\ref{cont v2-3}).


The algorithm presented here could also be used with Pad\'e approximants of higher degree or different funcional ansatz. However, we will not consider other ansatz due computational limitations.


\section{Conclusions}

\noindent

In this paper, it was shown an algorithm for solving nonlinear partial differential equations based on Pad\'e approximants. The algorithm was applied to the $\lambda\phi^4$ model in 4 dimensions by using two funcional ansatzes and it lead us to new solutions for the model. There are many recent papers proposing methods for solving differential equations \cite{Parkes,Malfliet,Kudryashov01,kudryashov,Parkes2,Vitanov1,Vitanov2,Vitanov3,Biswas1,Biswas2,He,Wu,Ma1,Ma2,Wang,Ebadi,Lee,Ruy} and the approach presented here could be an easier algorithm for applying to more complicated model.

\section*{Acknowledgments}

\noindent
I am thankful to H. Aratyn, J. F. Gomes and A. H. Zimerman for discussions. The author also thanks  FAPESP (2010/18110-9) for financial support.

\end{document}